# Spatiotemporal singular value decomposition for denoising in photoacoustic imaging with low-energy excitation light source

MENGJIE SHI, TOM VERCAUTEREN, AND WENFENG XIA*

*School of Biomedical Engineering & Imaging Sciences, King's College London, St Thomas' Hospital, London, SE1 7EH, UK*
*wenfeng.xia@kcl.ac.uk*

**Abstract:** Photoacoustic (PA) imaging is an emerging hybrid imaging modality that combines rich optical spectroscopic contrast and high ultrasonic resolution and thus holds tremendous promise for a wide range of pre-clinical and clinical applications. Compact and affordable light sources such as light-emitting diodes (LEDs) and laser diodes (LDs) are promising alternatives to bulky and expensive solid-state laser systems that are commonly used as PA light sources. These could accelerate the clinical translation of PA technology. However, PA signals generated with these light sources are readily degraded by noise due to the low optical fluence, leading to decreased signal-to-noise ratio (SNR) in PA images. In this work, a spatiotemporal singular value decomposition (SVD) based PA denoising method was investigated for these light sources that usually have low fluence and high repetition rates. The proposed method leverages both spatial and temporal correlations between radiofrequency (RF) data frames. Validation was performed on simulations and *in vivo* PA data acquired from human fingers (2D) and forearm (3D) using a LED-based system. Spatiotemporal SVD greatly enhanced the PA signals of blood vessels corrupted by noise while preserving a high temporal resolution to slow motions, improving the SNR of *in vivo* PA images by 1.1, 0.7, and 1.9 times compared to single frame-based wavelet denoising, averaging across 200 frames, and single frame without denoising, respectively. The proposed method demonstrated a processing time of around 50 $\mu s$ per frame with SVD acceleration and GPU. Thus, spatiotemporal SVD is well suited to PA imaging systems with low-energy excitation light sources for real-time *in vivo* applications.

## 1. Introduction

PA imaging is a promising hybrid biomedical imaging technique capable of spatially resolving spectroscopic contrast of tissue at high ultrasonic resolution and depths. It can provide structural, molecular, and functional information of biological tissues, and has shown great potential in various pre-clinical and clinical applications [1]. However, PA signals could be corrupted by electrical noise due to thermal effects, external hardware, and environmental interferences [2]. System-related noise (parasitic noise) is usually associated with solid-state laser systems with high-voltage Q-switching, while thermal noise is not restricted to PA excitation sources but is randomly distributed in the background.

Various algorithms have been proposed for PA signal denoising on single RF acquisition, including frequency filtering, wavelet transform, and signal decomposition. Frequency filtering using bandpass filters is straightforward to implement but is less efficient when noise and signal spectra significantly overlap [3]. Wavelet-based denoising can be effective, but the selection of the number of decompositions and the thresholding criteria are challenging to automate [4–6].

Inherent characteristics of PA signals of tissue can be exploited for PA denoising via different decomposition models. Empirical mode decomposition (EMD) could decompose PA signals into several intrinsic mode functions (IMFs), allowing for noise removal by filtering out noisy IMFs. But the denoising performance is quite limited by its dependency on selecting IMFs of noise [7–9]. SVD could also be used to remove noise and extract signals of tissue for PA imaging. Hill *et al.* [10] proposed an SVD-based PA denoising approach for reducing laser-induced noise in the signal domain. Haq *et al.* [11] investigated K-means singular value decomposition (K-SVD) to compute a sparse representation of PA images. The K-SVD based denoising method involved an iterative process for selecting SVCs corresponding to noise, requiring high computation cost and sources. SVD-based denoising techniques are promising for outperforming frequency filtering and wavelet-based methods but require proper subspace rank selection for the optimality of noise rejection.

Recently, laser diodes (LDs) and light-emitting diodes (LEDs) have shown promise as an alternative to high-power lasers owing to their compact size, safety, and low cost [12], and could be useful in several clinical applications such as the detection of joint inflammation and guidance of minimally invasive procedures [12–15]. Despite much lower optical fluence leading to poor SNRs, LDs and LEDs excitation sources have higher pulse repetition frequencies (several kHz) compared to conventional laser sources (10-100 Hz), which provides an opportunity for denoising by leveraging the spatial and temporal features of the signals. Anas *et al.* [16] proposed a recurrent network and a convolutional neural work to exploit the spatial and temporal dependencies respectively. Although they reported a considerable improvement over signal averaging using *in vivo* data, the performance could be limited by the use of averaged signals as the ground truths for training that could be affected by motion artefacts. Wang *et al.* [17] proposed a spatiotemporal image reconstruction method for dynamic photoacoustic computational tomography (PACT) and reported improved image quality and computational cost against conventional frame-by-frame reconstruction methods.

Spatiotemporal SVD has been used for ultrafast and super-resolution ultrasound (US) imaging for extracting blood clutters from tissue background with good sensitivity [18,19]. In US imaging, in contrast to tissue backscatter signals, slow blood flow signals have lower spatial coherence and power, so they could be separated in the singular value domain. A recent study by Al Mukaddim *et al.* [20] used spatiotemporal SVD in PA imaging to extract the dynamic cardiac signals during a cardiac cycle.

In this work, for the first time to our knowledge, we propose a spatiotemporal SVD for denoising in PA imaging dedicated to systems with low-fluence light sources. Different from [20], in this study, the background noise is considered dynamic (randomly distributed and having a low spatial and temporal coherence) while tissue signals are considered quasi-static (high spatial and temporal coherence). We validated our method on simulated data, and *in vivo* data acquired from a human volunteer using a LED-based system in comparison with averaging across multiple imaging frames, and wavelet denoising based on single frames.

## 2. Materials and methods

### 2.1 Randomized Spatiotemporal Singular Value Decomposition (rsSVD)

Conventional implementation of the SVD-based clutter filtering for US imaging [18] requires the formulation of a spatiotemporal Casorati matrix $S$ of dimension $(n_x \times n_z, n_t)$ based on an ultrasound data set with $n_t$ frames of 2D data of size $(n_x, n_z)$. SVD is then performed on the matrix $S$:

$$S = U\Delta V^* \quad (1)$$

where $U$ and $V$ correspond to the spatial and temporal singular vectors of $S$, respectively. $\Delta$ is a diagonal matrix with diagonal entries listed in descending order. Signals with higher energy and higher spatiotemporal coherence will be captured mostly by the largest singular vectors, which can be filtered out by multiplying a filtering matrix $I^F$. This produces the filtered signals as:

$$S^F = U\Delta I^F V^* \quad (2)$$

Implementing the full SVD of the matrix $S$ is associated with a high computational complexity $O\ (n_z \times n_x \times n_t^2)$, therefore leading to long computation times. Randomized SVD (rSVD) was proposed for accelerating the SVD process by low-rank matrix approximations [21] and could be implemented on GPU with PyTorch. rSVD is suitable for low energy-based PA denoising since tissue signals with intrinsically higher spatial and temporal coherence than noise resides in the first largest singular values. Therefore, a truncated SVD such as rSVD is practically efficient for estimating tissue subspace. The detailed mathematics and validation of rSVD using US images can be found in [22] and [23]. Here, we describe the critical steps for implementing randomized spatiotemporal singular value decomposition (rsSVD) for PA denoising.

A 3D RF matrix with a size of $(n_x, n_z, n_t)$ was formed using stacks of RF data obtained for each pulse of LED excitation with a dimension of $(n_x, n_z)$, where $n_x$ and $n_z$ denotes the number of transducer elements and the number of time steps, respectively. $n_t$ corresponds to the number of frames. The 3D RF matrix was then reshaped to the Casorati matrix, $S$, with a dimension of $(n_x \times n_z, n_t)$. Instead of calculating a full SVD of the matrix $S$, rsSVD generated a matrix $Q$ with rank $k$ as an approximation of the matrix $S$. We observed that the quasi-static tissue signals primarily corresponded to the largest $k$ rank singular values with $k \ll t$. To calculate $Q$, the matrix $S$ was first multiplied by a random matrix $R$.

$$S' = SR \qquad (3)$$

Each entry of $R$ was generated from a Gaussian distribution $N(0,1)$. The matrix $Q$ was obtained by performing the pivoted QR decomposition on the matrix $S'$.

$$Q = qr(S') \qquad (4)$$

Power iteration [21] was used to accelerate the convergence of the singular values to the full SVD results. Thus, the rank $k$ matrix $S$ was approximated as

$$S \approx QQ^*S \qquad (5)$$

where $Q^*$ represents complex conjugate transpose. The denoised PA data was obtained by the signal matrix $P$ as:

$$P = QQ^*S \qquad (6)$$

*2.2 Rank estimation based on singular vectors*

Thresholding strategies for SVD-based clutter suppression methods have been investigated in US imaging [18,24–26]. Estimators based on the characteristics of temporal and spatial singular vectors proposed in [26] were proven efficient for clutter filtering with US images. For the estimator based on the temporal singular vectors, Fig.1 (a) shows the double-side power spectral density (PSD) of all the temporal vectors for an *in vivo* RF acquisition on human fingers (see Sec. 2.4 below for details on the data acquisition). For each temporal vector, the bandwidth containing 99 % of the energy was computed at each side. Compared to noise signals, tissue signals had a narrower bandwidth which was associated with the temporal vectors of high correlation order [27]. A non-parametric estimator was therefore defined by the inflexion point where the 99 % bandwidth was expanded towards the entire frequency band. In contrast, the estimator based on the spatial singular vectors leveraged the spatial features within the tissue and noise subspaces. In Eq. (1), $|u_k|$ representing the intensity of the $k_{th}$ column of $U$ was observed to be highly correlated only within the tissue subspace. The correlation matrix of $|u_k|$ $(k = 1 \ldots n_t)$ could intrinsically reveal the differences in spatial statistics, indicating the boundary between the tissue and noise subspaces. The correlation matrix was referred as the spatial similarity matrix and the calculation was given in Eq. (7). Fig.1 (b) demonstrates the spatial similarity matrix acquired from the *in vivo* measurements of human fingers. The correlation square formed by the spatial vectors of high orders revealed the tissue subspace. As the thermal noise mostly contributed to the noise signals, the corresponding spatial vectors rendered uncorrelated due to the spatial randomness of the noise. Therefore, the rank estimator based on the spatial vectors was defined by the identification of the boundary of the correlation square.

$$C(n,m) = \frac{1}{n_x*n_z}\sum_{p}^{n_x*n_z} \frac{(|u\_n(p)|-\overline{|u\_n|})*(|u\_n(p)|-\overline{|u\_n|})}{\sigma\_n*\sigma\_m} \quad (m,n) \in [\![1,n\_t]\!]^2 \qquad (7)$$

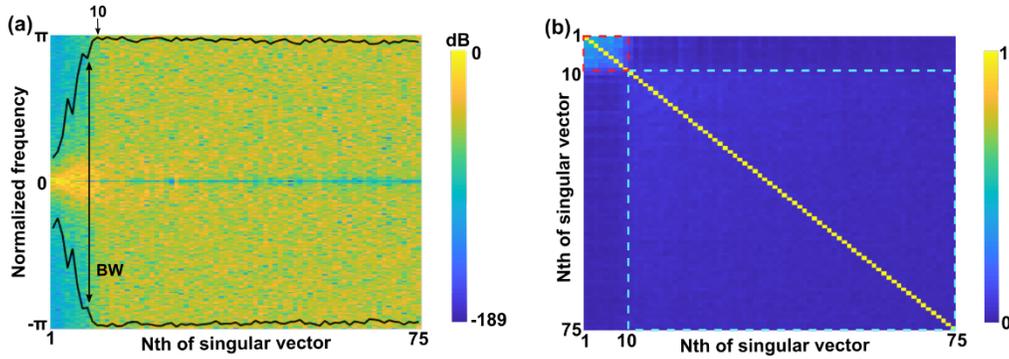
Fig. 1 Rank estimators based on temporal singular vectors (a) and spatial singular vectors (b) for an *in vivo* RF acquisition using human fingers. (a) Power spectral density (PSD) and 99% bandwidth (BW) of temporal singular vectors. (b) Correlation matrix of spatial singular vectors. The back arrow in (a) denotes the first inflexion point of the BW curve. The red dashed square from 1 to 10 represents the tissue subspace while the blue dashed square represents the noise subspace. Estimations were shown on singular vectors ranging from 0 to 75.

*2.3 Numerical simulations*

The spatiotemporal SVD based denoiser was first evaluated on simulated PA images. To generate the dataset, initial pressure distributions were created by simulating the optical fluence distributions on artificial vasculature using a Monte Carlo simulation program MCML [28]. A square region 40.0 mm (X) * 40.0 mm (Z) with the grid size of 0.1 mm was considered. The vascular pattern was acquired from the DRIVE database [29]. The optical properties of the vessels were specified by referring to those for oxygenated whole blood (150 g hemoglobin/liter) [30]. The optical scattering coefficient, the optical absorption coefficient, and the anisotropy of scattering at the wavelength of 850 nm was 0.5 $mm^{-1}$, 5.9 $mm^{-1}$, and 0.9 respectively. The background was assigned as a standard homogenous tissue with a uniform refractive index of 1.4, optical scattering coefficient of 10 $mm^{-1}$, optical absorption coefficient of 1.5 $mm^{-1}$, and anisotropy of 0.9. The PA excitation light was simulated as a homogeneous 38.4 mm photo beam delivering light at the tissue surface. The fluence distribution map was generated with 100 000 photons in 5 min.

A linear array US transducer with 128 sensor elements over a length of 38.4 mm was simulated in k-Wave [30]. Each element had a central frequency of 7 MHz and -6 dB bandwidth of 80.9 % for simulating the US sensors employed in the experimental setup (See section 2.4 below for details). During the forward simulations, PA waves were generated and propagated based on the initial pressure distribution map and then received by the sensor elements. To simulate RF acquisitions under a steady condition, the vessels spanning from 10.0 mm to 20.0 mm in the Z direction were assumed to be stationary over 100 frames (~1 s acquisition time using AcousticX). Furthermore, considering the slow movements during handheld imaging and mechanical scanning, the total 100 acquisitions were divided into 4 groups where the vessels region was sequentially moved along the Z direction with 0 mm, 1 mm, 2 mm, and -1 mm, respectively. Gaussian Random noise was added into the Casorati matrix of the simulated RF data. Three typical noise levels at -5 dB, -10 dB, and -15 dB in SNR were considered according to the noise distributions measured in vivo.

*2.4 PA imaging of the human fingers and forearm in vivo*

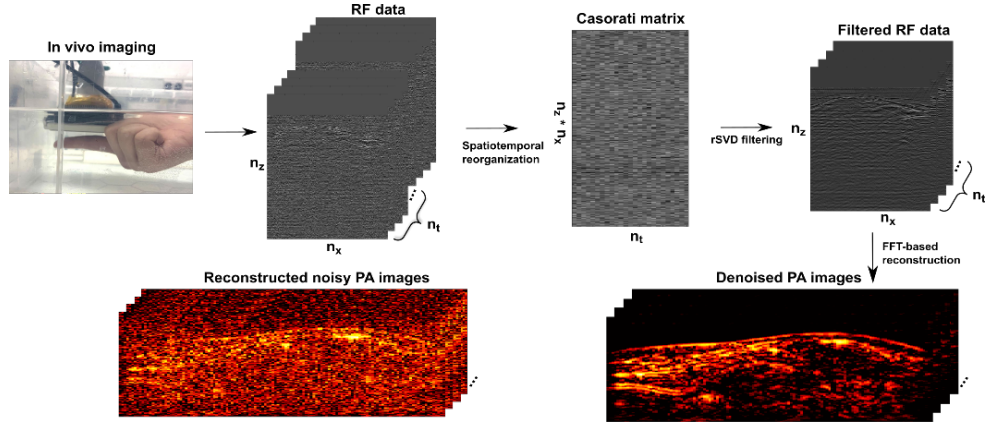
Fig. 2 Schematic illustration of spatiotemporal SVD based PA denoising.

Fig. 2 shows the spatiotemporal SVD based PA denoising framework with *in vivo* data. Human volunteer experiments were approved by the King's College London Research Ethics Committee (study reference: HR-18/19-8881). An LED-based PA/US imaging system (AcousticX, CYBERDYNE INC, Tsukuba, Japan) was employed for collecting *in vivo* data for validation [15]. US detection used an integrated PA/US probe consisting of a 7 MHz linear array transducer with 128 elements and two 850 nm LED arrays affixed on both sides.

To validate the proposed method for *in vivo* applications, especially superficial vasculature, fingers of a healthy volunteer were imaged with the imaging probe handheld underwater. Imaging was performed for around 22 s with 1536 frames of raw PA data captured for offline processing. Another validation experiment involved *in vivo* 3-D scanning of superficial vessels of the human forearm (see Fig. 6). The imaging probe was affixed on a motorised linear translation stage and moved in the Y-direction at a constant speed. The probe was translated for around 40 mm in 13 s along the Y-direction with 1299 frames of raw PA data acquired, which were then fed into a Fourier domain algorithm for image reconstruction [31]. The proposed denoising method was further compared with discrete wavelet transform (DWT). DWT was implemented on single frame basis by employing 'db4' as the mother wavelet and soft thresholding via universal thresholding [32].

### 2.5 Quantitative analysis

Reference metrics including peak-to-noise ratio (PSNR), edge preservation index (EPI) [33,34], and structure similarity index measure (SSIM) [35] were chosen for analysing the denoising performance on simulated data. With a denoised PA image $X$ and its corresponding noise-free image $Y$ of size $M \times N$, the PSNR is given by:

$$PSNR = 10 log_{10} \left( \frac{max(X)}{\frac{1}{MN}\sum_{i=1}^{M}\sum_{j=1}^{N}[X(i,j)-Y(i,j)]^2} \right) \qquad (8)$$

The simplified SSIM is computed as:

$$SSIM = \frac{(2\mu_x\mu_y+C_1)(2\sigma_{xy}+C_2)}{(\mu_x^2+\mu_y^2+C_1)(\sigma_x^2+\sigma_y^2+C_2)} \qquad (9)$$

where $\mu_x, \mu_y, \sigma_x, \sigma_y, \sigma_{xy}$ represents the means, standard deviations, and cross-variance for the denoised image $X$ and noise-free image $Y$, respectively. $C_1 = 0.01L^2$ and $C_2 = 0.03L^2$ where $L$ is 255 that denotes the dynamic range of the input images of data type uint8. The EPI is commonly used index for edge quality analysis especially during image denoising [34]. Eq. (10) defines the computation of the EPI.

$$EPI = \frac{\Gamma(\Delta s - \overline{\Delta s}, \ \widehat{\Delta s}-\overline{\widehat{\Delta s}})}{\sqrt{\Gamma(\Delta s - \overline{\Delta s}, \Delta s - \overline{\Delta s})\Gamma(\widehat{\Delta s}-\overline{\widehat{\Delta s}}, \widehat{\Delta s}-\overline{\widehat{\Delta s}})}} \qquad (10)$$

$$\Gamma(s_1, s_2) = \sum_{(i,j)\in ROI} s_1(i,j)s_2(i,j) \tag{11}$$

Where $\overline{\Delta s}$ and $\overline{\widehat{\Delta s}}$ represent mean values of the high pass Laplacian filtered ROIs $\Delta s$ and $\widehat{\Delta s}$ from the denoised image $X$ and noise-free image $Y$, respectively. In this study, EPI was calculated using the whole images $X$ and $Y$ instead of choosing ROIs explicitly.

To quantitatively evaluate the denoising performance in terms of image quality and sensitivity to slow motions using *in vivo* data, axial resolution and non-reference metrics including signal-to-noise ratio (SNR) and contrast-to-noise ratio (CNR) were measured.

For *in vivo* imaging with human fingers, strong PA signals were observed from a double layer structure that may correspond to a digital artery considering the apparent pulsations in real-time display. To further quantify the motion-induced blurring effect with frame averaging and comparing the sensitivity of spatiotemporal SVD to subtle movements, axial resolution was calculated at the vessel edges. An edge-spread function was obtained by drawing an axial profile across the two-layers structure (blue line in Fig. 5b). The full width at half maximum (FWHM) value was then measured as the axial resolution of the system.

Signals that may have corresponded to blood vessels and background noise were extracted from specified regions of interest (ROIs) of equal pixels based on reconstructed PA images. SNR was defined as the ratio of the mean amplitude of signal regions and standard deviation of background noise in decibels. CNR was given by the difference of mean amplitude in signal and noise regions divided by the standard deviations in the noise regions. They can be expressed as:

$$SNR = db\left(\frac{\frac{1}{n}\sum_{i=1}^{n}\mu_i}{\frac{1}{m}\sum_{j=1}^{m}\sigma_j}\right) \tag{12}$$

$$CNR = \frac{\left|\frac{1}{n}\sum_{i=1}^{n}\mu_i - \frac{1}{m}\sum_{j=1}^{m}\mu_j\right|}{\frac{1}{m}\sum_{j=1}^{m}\sigma_j} \tag{13}$$

where $\mu_i$, $\mu_j$ and $\sigma_b$ denote the mean signal amplitude, mean noise amplitude, and standard deviation of background noise for the $i^{th}$ and $j^{th}$ ROI of signals and noise, respectively.

## 3. Results

### 3.1 In silico experiments

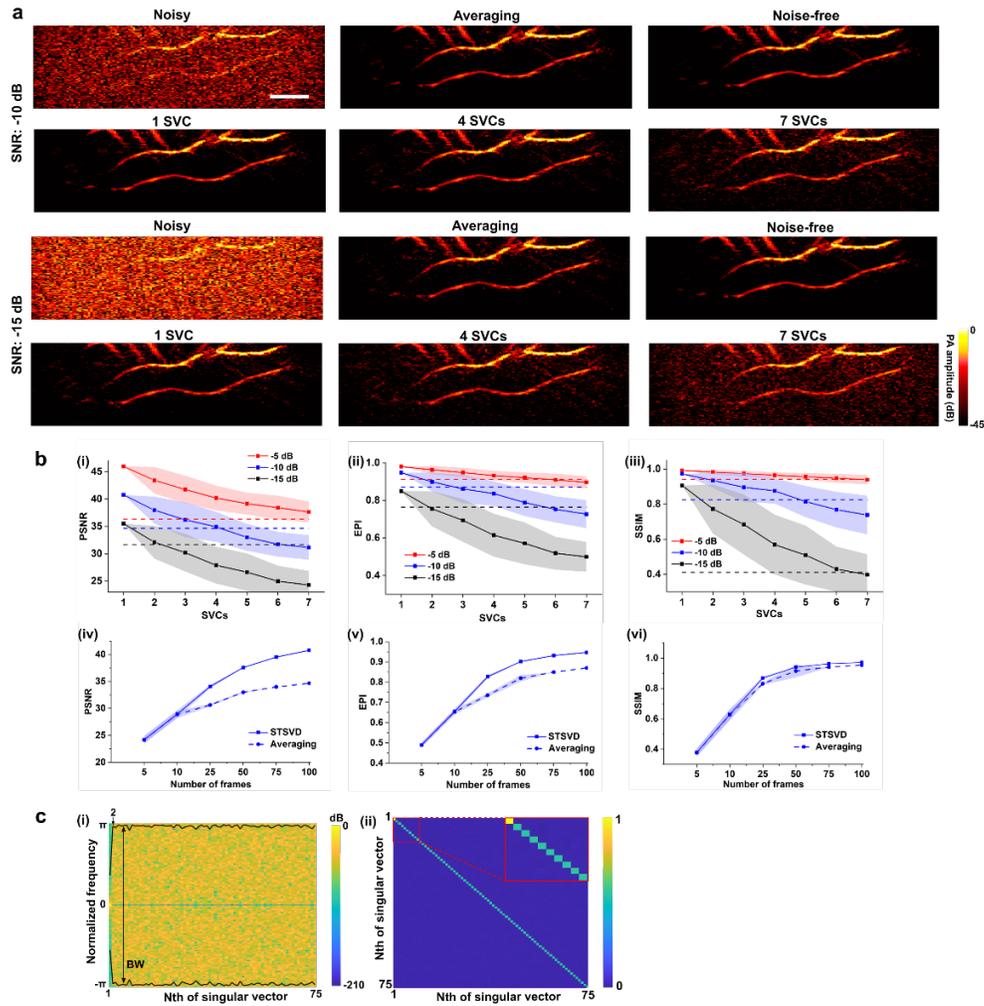

**Fig. 3** In silico comparison of spatiotemporal SVD (STSVD) and frame averaging on simulated vascular data without slow movements. **a.** Noisy and denoised photoacoustic (PA) images by STSVD and corresponding averaged and noise-free images. Two typical noise levels at -10 dB (i) and -15 dB (ii) respectively were compared. **b.** Quantitative analysis of STSVD and frame averaging in terms of peak-to-noise ratio (PSNR), edge preservation index (EPI), and structure similarity index measure (SSIM). (i)-(iii) reported the quantitative results of STSVD against different numbers of singular value components (SVCs) for signal reconstruction and frame averaging (dotted lines) using simulated vascular data at three noisy levels; 75 frames were used both for STSVD and averaging. (iv)-(vi) showed the quantitative results of STSVD and frame averaging against different number of frames in process using simulated vascular data at the middle noisy level (-10 dB in SNR). Data of STSVD in (i)-(vi) and averaging in (iv)-(vi) represent average values with shades denoting standard deviations. **c.** Temporal (i) and spatial criteria (ii) for rank selection for simulated vascular data without motions. All PA images share the same scale bar of 5 mm.

Fig. 3a shows denoising performance of averaging and spatiotemporal SVD using the simulated vascular data without slow movements. Two representative noise levels with the SNRs of -10 dB and -15 dB were chosen. Averaging over 100 frames reduced most of random noise in the background as well as enhanced the vessel signals. Visually speaking, spatiotemporal SVD achieved similar improvements in noise reduction and signal preservation with the first singular component. The denoising performance decreased with more singular value components (SVCs) used for the inversion. Considering the noisier data, spatiotemporal SVD had a comparable performance to averaging with the largest singular value.

Fig. 3b further demonstrates the impact of the number of SVCs (i - iii) and RF frames (iv - vi) on denoising. The performance by spatiotemporal SVD and averaging was quantified by

PSNR, EPI, and SSIM with noise-free images as reference. Significant decreases of the measured metrics were observed with the largest 7 SVCs. For all three noise levels, with the largest singular component, spatiotemporal SVD achieved higher PSNR, EPI, and SSIM than averaging. Fig. 3b (iv - vi) show the metrics performance against different number of RF frames in use. Statistical difference at a small number of frames (e.g., 5 or 10) were barely observed for spatiotemporal SVD and averaging. However, the performances were improved by STSVD with more frames in use. For example, PSNR, EPI, and SSIM of spatiotemporal SVD were 1.2-, 1.1-, 1.0-fold higher than those of averaging with 75 frames in process. Noted that both for spatiotemporal SVD and averaging, the enhancements in PSNR, EPI, and SSIM became less significant as the number of frames larger than 50.

The proposed rank estimators based on spatial and temporal vectors were applied on the simulated data. Fig. 3b (vii) and (viii) show the acquired PSD, 99% BW, and spatial similarity matrix. For the simulated data without motions, the tissue subspace was clearly identified by the first transition point during the evolution of the 99% bandwidth and the first correlation square. Those corresponded well to the experimental results with different numbers of SVCs in Fig. 3b (i) – (iii).

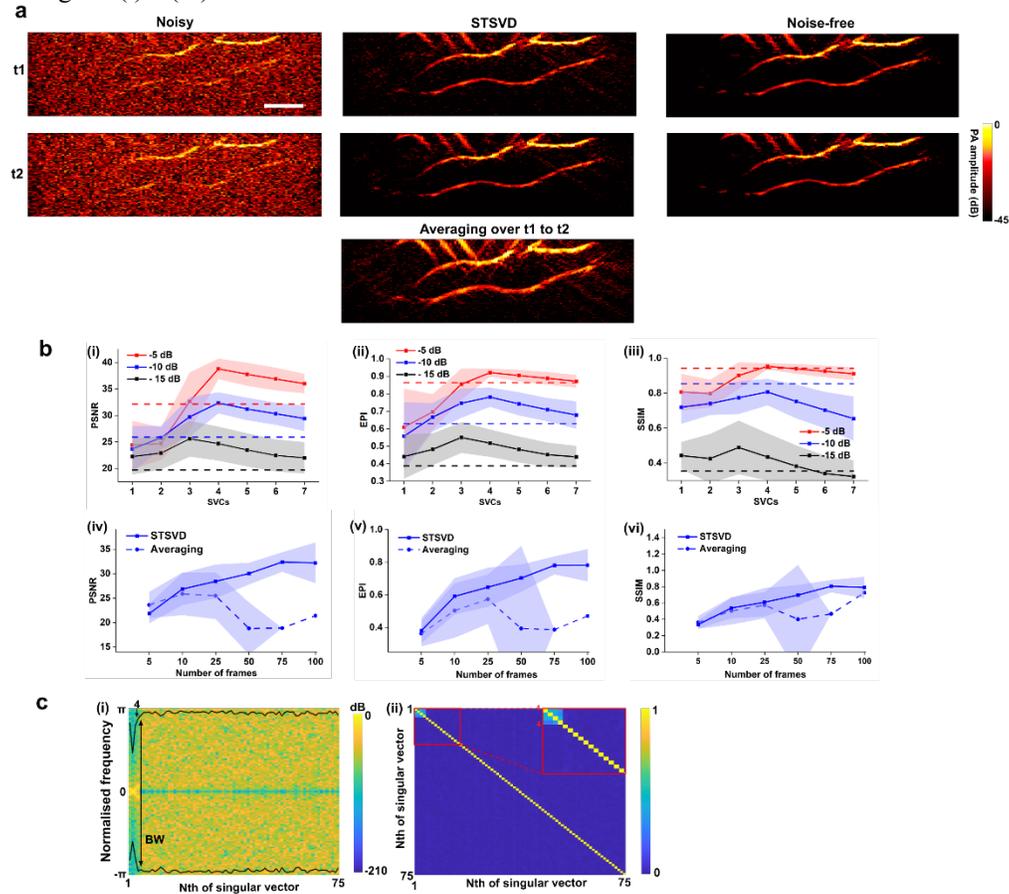

**Fig. 4** In silico comparison of spatiotemporal SVD (STSVD) and frame averaging on simulated vascular data with slow movements. **a.** Noisy and denoised photoacoustic (PA) images at two time points t1 and t2 by STSVD and corresponding averaged and noise-free images; noise level: -10 dB in SNR. **b.** Quantitative analysis of STSVD and frame averaging in terms of peak-to-noise ratio (PSNR), edge preservation index (EPI), and structure similarity index measure (SSIM). (i)-(iii) reported the quantitative results of STSVD against different numbers of singular value components (SVCs) for signal reconstruction and frame averaging (dotted lines) using simulated vascular data at three noisy levels; 75 frames were used both for STSVD and averaging. (iv)-(vi) showed the quantitative results of STSVD and frame averaging against different number of frames in process using simulated vascular data at the middle noisy level (-10 dB in SNR). Data of STSVD in (i)-(vi) and averaging in (iv)-(vi) represent average values with shades

denoting standard deviations. **c.** Temporal (i) and spatial criteria (ii) for rank selection for simulated vascular data with motions. All PA images share the same scale bar of 5 mm.

Fig. 4 further compares the denoising performance of averaging and spatiotemporal SVD in terms of signal sensitivity by using simulated data with slow motions. Two frames (corresponding to moment t1 and t2) were chosen from the total of 100 frames for comparison. The time interval was carefully selected to include a certain number of synthetic motions. As shown in Fig. 4a, averaging over t1 to t2 effectively suppressed background noise, but rendered motion artefacts indicated by the blurring vessel edges. In contrast, spatiotemporal SVD resolved the sharp edges of the blood vessels and removed noise in the background. Fig. 4b explicitly demonstrates its statistical performance regarding the optimisation of SVCs and frames. The optimal performance of spatiotemporal SVD shown in Fig. 4a was achieved with the first 4 SVCs and 75 RF frames. However, for frame averaging, it is worth noting that the improvement of denoising performance was not consistent with the increase of the processed number of frames due to the integrated slow motions. In Fig. 4b (iv) – (vi), the PSNR, EPI, and SSIM (blue dashed lines) suddenly decreased as the number of frames increased from 25 to 50.

The non-parametric rank estimators were also evaluated with the simulated data containing slow motions (Fig. 4c). The bandwidth and spatial similarity matrix both gave an optimal threshold around the first 4 SVCs. The estimated threshold also appeared consistent with the results acquired by individually testing possible numbers of SVCs (Fig. 4b (i) - (iii)).

### *3.2 In vivo experiments with human fingers*

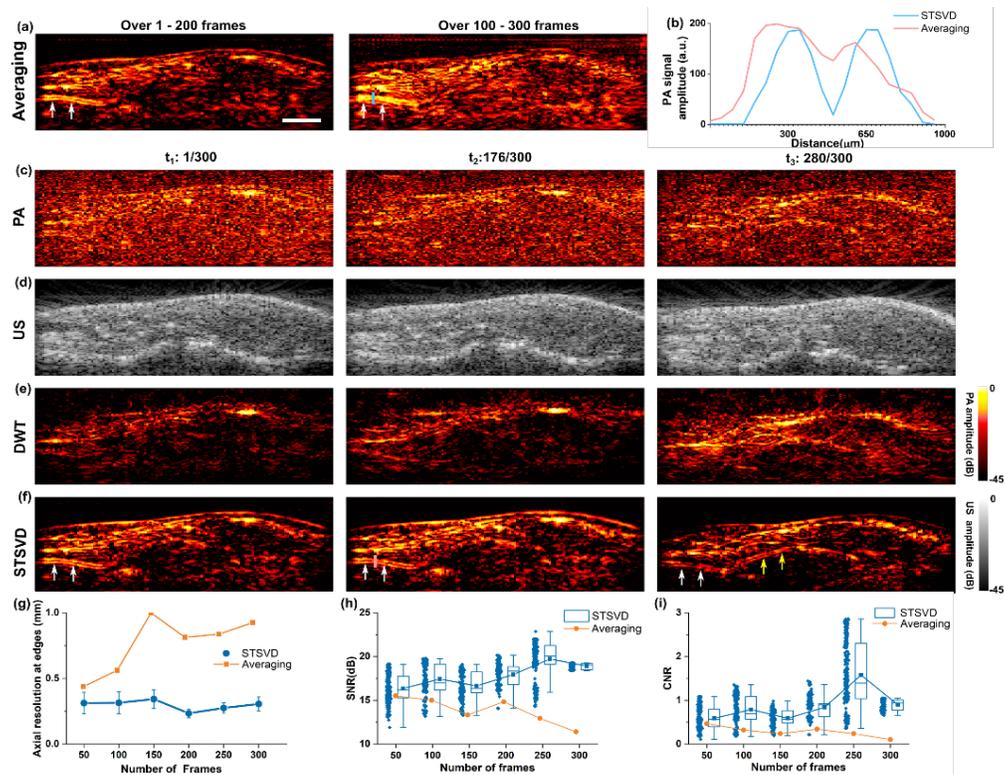

**Fig. 5.** In vivo comparison of (a) PA frame averaging on human finger data with (c) noisy PA signal, (d) corresponding US signal, (e) discrete wavelet transform (DWT) PA denoising, and (f) spatiotemporal SVD (STSVD) PA denoising. (b) Axial profiles drawn along the lines marked in the denoised photoacoustic (PA) image by spatiotemporal SVD and corresponding averaged image. (g)-(i) Quantitative performance of STSVD and frame averaging in terms of axial resolution at the vessel edges, signal-to-noise ratio (SNR), and contrast-to-noise ratio (CNR). Data of STSVD in (g) represent average values with error bars denoting standard deviations. The box charts of STSVD in (h) and (i) represent the summary statistics. The box is determined by the 25[th] and 75[th] percentiles while the whiskers are determined by the

5th and 95th percentiles. Mean and median values are represented. Mean values were connected by lines. All PA and ultrasound (US) images share the same scale bar of 5 mm.

Fig. 5 compares representative results of denoised PA images through frame averaging, DWT, and spatiotemporal SVD using *in vivo* data from human fingers. Differing from averaging and spatiotemporal SVD, the wavelet signal denoiser was employed on a frame-by-frame basis. Quantitative results in terms of axial resolution at vessel edges, SNR, and CNR with averaging and spatiotemporal SVD were reported against the different number of frames in use. A total number of 300 frames with an acquisition time of ~5 s was processed (see. Supplementary materials for real time display). Spatiotemporal SVD employed 200 frames (acquisition time: ~3 s) in the Casorati matrix considering the optimality of noise suppression and temporal resolution. Comparisons were performed between noisy PA images, spatiotemporal SVD denoised PA images at three different time points $t_1, t_2, t_3$, and frame averaged results during two-time intervals.

Before denoising, PA signals from blood vessels were buried by randomly distributed background noise of high intensity, leading to poor visibility of vascular structures, and limited tissue penetration depth. The wavelet-based denoiser reduced the noise but also generated substantial signal loss with a mean SNR of 9.70 dB and CNR of 0.13 over 200 frames. Spatiotemporal SVD greatly suppressed the background noise and maintained high sensitivity to fine tissue structures. e.g., the improved contrast of a digital artery denoted by white arrows in Fig. 5f with a two-layer structure. Averaging over the first 200 frames achieved an improved SNR of blood vessels, but motion artefacts were evident as indicated by the blurring edges (white arrows in Fig. 5a). Furthermore, movements could also be observed during $t_2$ to $t_3$ where additional vascular signals became visible (yellow arrows in Fig. 5f). As expected, the reconstructed image using the averaged RF data of this interval was largely contaminated by motion artefacts as well as background noise. On the contrary, spatiotemporal SVD effectively reduced the background noise and had appreciable improvements in restoring fine structures for *in vivo* data containing slow motions. The axial resolution was 0.235 mm ± 0.07 for spatiotemporal SVD with 200 frames in decomposition, which was much smaller than that of averaging (0.81 mm) across the same frames. It is noticed that in Fig. 5h and Fig. 5i, SNR and CNR with frame averaging decreased with the number of averaged frames when it is larger than 200 due to the accumulated stationary noise and slow motions.

*3.3 In vivo 3-D scanning of human forearm*

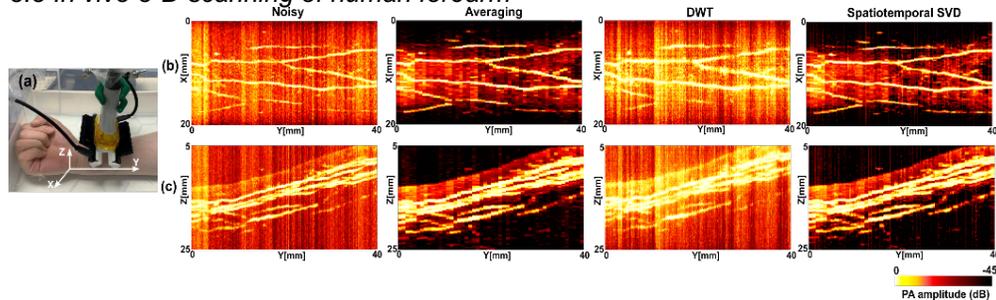

**Fig. 6.** Comparison of spatiotemporal SVD, discrete wavelet transform (DWT), and averaging for PA denoising with 3-D scanning of the human forearm *in vivo*. (a) The experimental set-up with the right forearm of a healthy volunteer imaged in water. (b, c) Maximum intensity projections (MIPs) of the reconstructed 3-D PA volumes onto XY-plane (b) and ZY-plane (c) with the depth Z ranging from 5 mm to 25 mm.

Fig. 6 shows the maximum intensity projections (MIPs) of 3D PA images of the human forearm within 5 to 25 mm in depth with and without denoising. The same number of frames (25 frames) were used for frame averaging and spatiotemporal SVD while the wavelet-based method was performed on a single frame basis. Superficial vasculature in the human forearm was greatly enhanced by averaging and spatiotemporal SVD. On the contrary, the wavelet denoiser failed to mitigate the noise and degraded the spatial resolution of vessels. Spatiotemporal SVD largely retained the microvasculature distribution in 3D. However, severe distortions including discontinuity and edge jags of blood vessels were shown by using signal averaging.

## 4. Discussion and conclusions

This study presented a denoising approach based on spatiotemporal SVD for low-fluence based PA imaging. Compared to signal averaging and wavelet transform, spatiotemporal SVD based denoiser explores both the spatial and temporal statistics of RF data. The proposed method was validated using simulated vascular data, *in vivo* 2D PA image data acquired from human fingers, and 3D scanning data of human forearm, outperforming the frame averaging and DWT in terms of suppression of background noise, sensitivity to slow motions, and vascular signal specificity. The axial resolution, SNR, and CNR values of SVD denoised images were 2.4, 0.7, 4.7 times higher than those of images with repetitive signal averaging. We demonstrated a processing time of about 10 ms for a matrix of size $600 \times 128 \times 200$ using a GPU (Tesla T4, Pytorch 1.11.0) [21], indicating the feasibility of real-time processing.

Low-cost light sources such as LEDs and LDs are promising for clinical translation. However, the imaging performance of low-fluence based PA systems is hindered by low output energy of these light sources, especially for deep tissue. Averaging over hundreds to thousands of time frames acquired with high pulse repetition frequencies (PRFs) is a common way to improve the image quality but decreases the frame rate, leading to an insufficient temporal resolution that is critical for real-time studies of physiological responses. For handheld imaging or 3D scanning, simple averaging will bring important motion artefacts, causing severe distortions.

Since the temporal resolution of low-fluence based PA imaging systems is sufficiently high, signals from tissue present high spatial and temporal correlation during high-speed acquisitions while thermal noise can reach its full randomness and appears to be spatiotemporally uncorrelated. Therefore, spatiotemporal fluctuations of signals and noise could be efficiently characterised by singular vectors. Besides, compared to conventional solid-state laser-based PA imaging systems, systems employing low-fluence light sources are unlikely to be affected by high-intensity parasitic noise. Thus, denoising was performed by preserving the low-order singular values contributed by signals from critical targets including deep depth with high spatiotemporal correlations. Validation results on simulated data showed that noise reduction with minimal degradation of critical signals could be achieved by constructing the signal subspace using the largest $k$ singular values ($k \ll t$). Besides, the rank estimators based on the characteristics of the singular components were applied on both simulated and *in vivo* data, showing efficient for automatically resolving the boundary of tissue and noise subspaces. The number of frames for STSVD was determined considering the temporal resolution and denoising performance. For systems with lower repetition frequencies, real-time denoising with STSVD could be implemented with a sliding window on time frames.

Spatiotemporal SVD was also validated with 3D scans of the human forearm *in vivo*. Variations of microvasculature between different frames were challenging to capture by using signal averaging. It is worth noting that the 3-D blood vasculature were well restored by spatiotemporal SVD with minor artefacts. However, outliers that were projected as a set of parallel lines could be observed, especially for initial frames of scanning, which could be attributed to unstable scanning conditions that interrupted the consistency of signal or noise statistical distributions.

Spatiotemporal SVD outperformed conventional single frame based denoiser such as DWT as it leveraged coherence information from multiple frames. State-of-the-art denoisers such as BM3D [36] also showed little improvement for realistic PA data in the image domain (see Supplementary Materials). Learning-based denoisers based on supervised models require a large noisy dataset with clear data as ground truth for training, which is challenging and expensive [37,38]. Self-supervised models were explored based on single noisy image [39,40], but the denoising performance could be improved with extra information of noise distribution.

Compared to averaging, spatiotemporal SVD could generate denoised *in vivo* PA data with higher fidelity that could be used as ground truth to develop learning-based denoising models in the future.

We conclude that spatiotemporal SVD is well suited for denoising in PA imaging with low-fluence light sources.

## 5. Funding

This research was funded in whole, or in part, by the Wellcome Trust [203148/Z/16/Z, WT101957, 203145Z/16/Z], the Engineering and Physical Science Research Council (EPSRC) (NS/A000027/1, NS/A000050/1, NS/A000049/1). For the purpose of open access, the author has applied a CC BY public copyright licence to any Author Accepted Manuscript version arising from this submission.

## 6. Disclosures

TV is co-founder and shareholder of Hypervision Surgical Ltd, London, UK. He is also a shareholder of Mauna Kea Technologies, Paris, France.

## 7. Data availability

Data underlying the results presented in this paper are not publicly available at this time but may be obtained from the authors upon reasonable request.